\documentclass[10pt]{article}
\usepackage[margin=1in]{geometry}                
\usepackage[parfill]{parskip}    
\usepackage{authblk}
\usepackage{caption}
\usepackage{subcaption}
\usepackage{graphicx}
\usepackage{float}
\usepackage[sort&compress,numbers]{natbib}
\usepackage{amssymb}
\usepackage{tensor}
\usepackage{xcolor}
 \usepackage[english]{babel}
\usepackage{epstopdf} 
\usepackage{amsmath}
 \numberwithin{equation}{section}
\usepackage{mathrsfs}
\usepackage{varioref}
\usepackage{hyperref} 
\usepackage{textcomp}
\newcommand{\appendixpage}

\hypersetup{
    pdfstartview={FitH},    
    colorlinks=true,       
    linkcolor=cyan,          
    citecolor=green,        
    filecolor=blue,      
    urlcolor=blue           
}

\DeclareFontFamily{OT1}{pzc}{}
\DeclareFontShape{OT1}{pzc}{m}{it}%
             {<-> s * [0.900] pzcmi7t}{}
\DeclareMathAlphabet{\mathscr}{OT1}{pzc}%
                                 {m}{it}

\begin{document}
\definecolor{orange}{rgb}{0.9,0.45,0}
\definecolor{applegreen}{rgb}{0.55, 0.71, 0.0}
\definecolor{blue}{rgb}{0.0,0.0,1.0}
\definecolor{red}{rgb}{1.0,0.0,1.0}
\newcommand{\spl}[1]{{\textcolor{orange}{[SP: #1]}}}

\newcommand{\co}[1]{{\textcolor{applegreen}{[comment: #1]}}}
\newcommand{\jh}[1]{{\textcolor{cyan}{#1}}}
\newcommand{\mc}[1]{{\textcolor{blue}{[#1]}}}
\newcommand{\quit}[1]{{\textcolor{red}{[#1]}}}

\title{Exploring the UV and IR of a type-II holographic superconductor using a dyonic black hole}
\author[1]{Manuel de la Cruz-López}
\author[1]{Jhony A. Herrera-Mendoza}
\author[1]{Roberto Cartas-Fuentevilla}
\author[1]{Alfredo Herrera-Aguilar}

\affil[1]{\begin{small}Instituto de Física, Benemérita Universidad Autónoma de Puebla, Apdo. Postal J-48, CP 72570, Puebla, México\end{small}}

\date{}                                           

\maketitle
{\it{Email:}}  
\href{mailto:mdelacruz@ifbuap.buap.mx}{mdelacruz@ifuap.buap.mx},  \
\href{mailto:jherrera@ifuap.buap.mx}{jherrera@ifuap.buap.mx},  \
\href{mailto:rcartas@ifuap.buap.mx}{rcartas@ifuap.buap.mx},  \
\href{malito:aherrera@ifbuap.buap.mx}{aherrera@ifuap.buap.mx},

\begin{abstract}

In this study, we investigate a type-II holographic superconductor with a perturbative scalar field over a (3 + 1)-dimensional electric and magnetically charged planar AdS black hole. After consistently decoupling the scalar field sector from the complete Einstein-Maxwell-Scalar system, we delve into the thermodynamical properties of the background relevant for the dual description of the Ginzburg-Landau density of superconducting states. 
The adoption of a London gauge allowed us to consider the magnetic field as a uniform external field over which the holographic superconductor is subject. This consideration enables a consistent description of the appearance of Abrikosov vortex lattices typical in type-II superconductors. 
Thus, by matching near horizon and boundary expansions of the scalar field, we obtained an expression for the upper critical magnetic field as a function of temperature in both, the canonical and grand canonical ensemble. These novel results confirm that our perturbative scalar field model consistently reproduces the well-known temperature behavior of the upper critical magnetic field according to the Ginzburg-Landau theory and other Abelian-Higgs holographic developments for type-II superconductors.
In addition, a new analysis of the scalar field equation in terms of a Schrödinger potential led us to observe the existence of potential wells distributed along the holographic coordinate. We interpret these regions with a local minimum as those in which bound states can exist, dual to the Cooper pairs density. These results provide evidence for the existence of an IR order parameter near the extremality. In view of this, we performed a closer inspection of the IR effective scalar equation in which the geometry adopts a Schwarzschild AdS$_{2}\times \mathbb{R}^{2}$ structure.


\end{abstract}

\section{Introduction and summary}
The celebrated Ginzburg-Landau theory is a powerful tool to explore order parameter phase transitions in Condensed Matter Systems. This theory is structured as a mean-field theory close to the transitions and as a function of a complex scalar field, coupled with dynamical gauge fields. Due to the spontaneous symmetry-breaking mechanism, the order parameter acquires a non-trivial vacuum expectation value (VEV) corresponding to the condensation of the electron density of Cooper pairs, defining the transition to a superconducting state. An emergent phenomenon (the Shubnikov state) was described by Abrikosov considering a wise assumption \cite{Abrikosov:1956sx}: from the linear approximation to the equations of motion and by symmetry arguments, he found the regime in which the magnetic field penetrates the superconducting sample, arranged in a lattice structure, extending the theory far beyond the London-Meissner regime of type-I superconductors\footnote{Nowadays, Type-II superconductors distinguish from Type-I ones at theoretical level due to Abrikosov.}. The fundamental cell of this lattice has a triangular structure with quantized magnetic fluxes (vortices) surrounded by superconducting currents. Using this solution on the free energy of the system, it is possible to prove the stability of the lattice configuration over the normal (conducting) state. If the magnetic field is strong enough, the superconducting state is destroyed and one open problem is to design experimentally materials with temperature/magnetic field ratio in such a way that the supercurrent persists at high temperatures. The latter question has a non-trivial answer due to the strong coupling nature of high-temperature superconductors, for which a theoretical description is needed far beyond the perturbation theory\footnote{The cuprates present high-temperature superconductivity, however, lack of complete theoretical description to describe them.}.

In this respect, one of the most promising theories is the AdS/CMT correspondence (whose ancestry comes from \cite{Maldacena:1997re}), in which Ginzburg-Landau Lagrangian is viewed as the effective boundary theory of gravitational systems. The corresponding bulk spacetime is a solution of Einstein-Maxwell-Scalar theories with black holes living in the deep interior, being the Hawking temperature, the thermodynamic temperature of the asymptotically AdS boundary theory. The boundary values of the Maxwell field give the chemical potential and the density of current of the dual system. AdS/CMT has many advantages at the operational level such as the use of the classical gravity regime to model strong coupling superconducting systems. Also, by standard interpretations of holography, the existence of a second-order transition from the state of vanishing scalar field to a new one in which the scalar has non-vanishing value, turning the background unstable to form scalar hair, was proved in \cite{Gubser:2008px}. The condensate of this scalar field is dual to the VEV of the superconducting order parameter \cite{Hartnoll:2008vx}. Within AdS/CMT we can compute the superconductor state (and superfluidity) correlation functions of gauge fields in curved spacetime, conductivity at linear response, as well the temperature dependency of order parameter from critical transition temperature to zero temperature; to name a few important developments. In all these quantities, holography has a good agreement concerning the qualitative behaviors of the temperature dependence of the dual system.

Holography considering neutral black holes in $3+1$ dimensions have been extensively studied since the seminal work \cite{Hartnoll:2008vx}. An important difference from charged black hole ones is that in the former, the Hawking temperature scales linearly with the horizon radius. Therefore, implementing the so-called decoupling limit, the system gives rise to the Abelian-Higss theory on Schwarzschild-AdS black hole background which is dual to Ginzburg-Landau mean-field theory with a global $U(1)$ conserved current. The effects of black hole rotational parameters developed by some of the current authors show that these parameters are traced in qualitative changes of the droplet and vortex solutions \cite{PhysRevD.106.L081902} (also see \cite{Nakano:2008xc} for the numerical proof of condensation in magnetically charged background). An essential aspect of the program of holographic superconductivity is the interaction between gauge fields and scalar fields. In \cite{Gubser:2008px} and \cite{Hertog:2006rr} it was proven that their interaction is crucial for hair formation (the scalar field mass acquires contribution from the $t$-component of the gauge field). A further development, considering the full backreaction between gravity and the Abelian-Higss sector, shows interesting physics: the hair formation arises independent of the charge $q$ of the scalar field. Moreover, the hair condensate can be sustained by the gravitational throat of the near horizon geometry which has topology AdS$_{2}\times \mathbb{R}^{2}$, even with a small electric charge $q$ (hence goes beyond the decoupling limit $q\mapsto\infty$) \cite{Hartnoll:2008kx}. Additionally, the interesting phenomena of Abrikosov lattice have been realized by using holographic setups, for instance, \cite{Montull:2009fe, Maeda:2009vf, Albash:2009iq, Adams:2012pj, Srivastav:2023qof, Xia:2019eje}.

Charged dyonic black holes have been known for a long time \cite{Romans:1991nq}. In this case, the dependency of temperature on the horizon radius becomes more intricate. A perturbative scalar field over the Einstein-Maxwell-AdS background constitutes one of the seeds of scalar hair formation driven by instabilities of the background \cite{Gubser:2008px}. In this system, there is no scalar contribution to the energy-momentum tensor in the Einstein equations, nevertheless, it has important novel developments, namely, it was used to study the Hall current \cite{Hartnoll:2007ai}, the low-frequency conductivity \cite{Horowitz:2009ij}, and the IR quantum criticality of the strange metals \cite{Faulkner:2009wj}; to name a few important contributions. Prominent for this work, is the analysis carried out in the second seminal work \cite{Hartnoll:2008kx}, in which the onset of holographic superconductivity (i.e., a small scalar field close to the transition temperature) is shown in the dyonic background exhibiting the formation of the non-trivial scalar field profile in the canonical ensemble. The near extremal limit, close to zero temperature, is analyzed in \cite{Albash:2008eh} demonstrating the existence of a droplet solution with Hermite polynomial behavior. In these latter works, the scalar field was considered purely real.

Concomitant with the previous lines, in this work we deal with an asymptotically AdS dyonic black hole with a planar horizon and then, consider the scalar field as a perturbation using a suitable decoupling limit from the Einstein-Maxwell-Scalar theory. The motivation for this choice is to use the so-called matching method to obtain the phase space of the superconductor, i.e., the curves of magnetic field versus temperature in both, the canonical and grand canonical ensemble.
In addition, similar to the holographic Abelian-Higgs model, a decoupled scalar field from Einstein-Maxwell theory, with functional dependency $\psi=\psi(x_{1},x_{2},r)$; being $r$ the holographic coordinate and $x_{i}$ the boundary theory ones, allow to design the periodicity properties of the solution to construct a lattice; as Abrikosov does\footnote{In this respect, in \cite{Donos:2020viz} the lattice was constructed with periodicity properties of boundary spacetime coordinates.
}

Since we are also interested in exploring the IR superconductor, we analyze the Schrödinger potential associated with the scalar field and prove the existence of bound states close to extremality. Furthermore, in this regime, in which Schwarzschild AdS$_{2}\times\mathbb{R}^{2}$ dictates the holographic IR description, we explore the behavior of the effective scalar field at near-zero temperatures, such as the asymptotic values and the conditions over the electromagnetic charges of the black hole triggering instabilities of the scalar field solution; representing a signal of IR hair formation. 
These results constitutes an additional step towards the general understanding of holographic $s$-wave bosonic superconductivity inaugurated by \cite{Gubser:2008px,Hartnoll:2008vx}.

\section{EMS setup}\label{Backreactionfull}

Consider the Einstein-Maxwell-Scalar theory (EMS) with a negative cosmological constant in $3+1$ dimensions
\begin{equation}\label{EMSaction}
    S=\dfrac{1}{\kappa_{4}^{2}}\int d^{4}x\sqrt{-g}\left[ \left( R+\dfrac{6}{L^{2}} \right)-\dfrac{L^{2}}{4}\epsilon_{1}^{2} F^{2}-L^{2}\epsilon_{2}^{2}\left( \vert D\psi\vert^{2}+m^{2}\vert\psi\vert^{2}\right) \right],
\end{equation}
where $\epsilon_{1}^{2}\equiv\kappa_{4}^{2}/g_{m}^{2}$ and $\epsilon_{2}^{2}\equiv\kappa_{4}^{2}/ g_{s}^{2}$ control the interaction between gauge and scalar fields with gravity, respectively. We are interested in spacetime metrics with AdS asymptotic and planar horizon topology
\begin{equation}\label{theMetric}
    ds^{2}=-U(r)dt^{2}+\dfrac{r^2}{L^2}\left(dx_{1}^{2}+dx_{2}^{2}\right)+\dfrac{dr^{2}}{U(r)},
\end{equation}
supported by electric and magnetic charges, i.e., a dyonic black hole. The Maxwell and the scalar field energy-momentum tensors
\begin{equation}\label{EMtensors}
\begin{array}{rcl}
T^{Max}_{ab}&\equiv& \dfrac{1}{2}F_{ad}F_{b}^{d}-\dfrac{1}{8}F^{2}g_{ab},\\ \\
T^{scalar}_{ab}&\equiv& D_{a}\psi D_{b}^{*}\psi^{*}-\dfrac{1}{2}\left(\vert D\psi\vert^{2}+m^{2}\vert\psi\vert^{2}\right)g_{ab},
\end{array}
\end{equation}
coupled with Einstein tensor via $\epsilon_{i}^{2}L^{2}$ defined above in (\ref{EMSaction}). Note that the choice of the potential for the scalar field is already fixed; knowing it effectively reproduces a large class of phenomenology in bottom-up constructions (as the Abelian-Higgs $s$-wave model \cite{Hartnoll:2008vx}). With (\ref{EMtensors}), the field equations read
\begin{subequations}\label{einsteineq2}
   \begin{align}
  R_{ab}&= \left(-\dfrac{3}{L^{2}}+\dfrac{\epsilon_{2}^{2}L^{2}}{2} m^{2}\vert\psi\vert^{2}-\dfrac{\epsilon_{1}^{2}L^{2}}{8} F^{2}\right)g_{ab}+\dfrac{\epsilon_{1}^{2}L^{2}}{2}F_{ad}F_{b}^{d}+\epsilon_{2}^{2}L^{2} D_{a}\psi D_{b}^{*}\psi^{*}\label{eins1} ,\\ \nonumber \\    
       \nabla^{a}F_{ab} &=\left(\dfrac{\epsilon_{2}}{\epsilon_{1}}\right)^{2}\left[ iq\left(\psi^{*}\nabla_{b}\psi-\psi\nabla_{b}\psi^{*}\right)+2q^{2}A_{b}\vert\psi\vert^{2}\right]\label{eins2},\\ \nonumber \\
       0 &= \nabla_{a}\nabla^{a}\psi-iq\left(2A^{a}\nabla_{a}+\nabla^{a}A_{a}\right)\psi-q^{2}A^{a}A_{a}\psi-m^{2}\psi\label{eins3}.      
\end{align}  
\end{subequations}
\subsection{Decoupling limit and the dyonic background}\label{decouplinS}

\begin{figure}[h]
    \centering
    \includegraphics[width=0.6\textwidth]{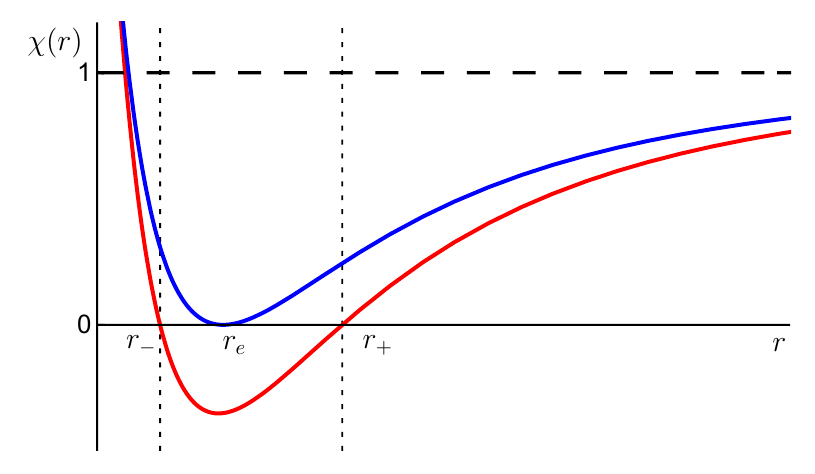}
    \caption{\begin{small}The blackening factor $\chi (r)$ potential (see (\ref{metricCoef})) exhibiting the two horizons and the extremal case labeled as $r_{e}$. The two horizons shrink into a single one at extremality condition (\ref{extremal2}). As we stated in the body of the text, violating the bound (\ref{extremalC}) translates into the fact that the real roots become complex as well, leaving the spacetime with a naked singularity. As a result of this analysis, $r_{+}$ represents the outermost horizon in the non-extremal case.\end{small}}
    \label{fig:outerh}
\end{figure}

The identification of $g_{s}$ with the charge $q$ of the scalar field and a proper re-scaling of the scalar and gauge fields $\psi\mapsto\psi/q$, $A\mapsto A/q$, enables us to consider the limit $q\mapsto\infty$ while $g_{m}^{2}\mapsto 0$. This implies that the effect of the scalar sector from the complete Einstein-Maxwell-Scalar system can be neglected. The following equations in this scalar decoupling limit arise\footnote{The decoupling limit of the Maxwell-Higgs sector, is accomplished by taking $A_{a}\mapsto A_{a}/q$, $\psi\mapsto\psi/q$, shearing the same coupling with gravity. The black hole that arises in this setup is the neutral Schwarzschild-AdS one.}
\begin{equation}\label{EinsMax1}
\begin{array}{rcl}
     G_{ab} &=& \dfrac{3}{L^2}g_{ab}+L^{2}\epsilon_{1}^{2}T^{Max}_{ab},\\ \\
     \nabla^{a}F_{ab} &=& 0,
\end{array}
\end{equation}
whereas the decoupled scalar field equation (\ref{eins3}) remains essentially unchanged. In this decoupling limit, the scalar field is considered as a perturbation over the dyonic black hole in contrast with the Abelian-Higgs theory. Nevertheless, following the arguments of the seminal works \cite{Hartnoll:2008kx} and \cite{Albash:2008eh}, in this report, we extend the primary constructions with the aid of a complex scalar field with a nontrivial phase to get the vortex structure; leaving the build of a lattice array for further works. To achieve this and exploit the $U(1)$ gauge invariance of Electrodynamics, we use the \emph{London gauge} for the magnetic component
\begin{equation}\label{londongauge}
    A_{2}=Bx_{1}dx_{2}, \qquad A_{t}=A_{t}(r)dt.
\end{equation}
The Einstein-Maxwell equations (\ref{EinsMax1}) that entail from the above ansatz read 
\begin{subequations}\label{eins0}
   \begin{align}
    A_{t}''+\dfrac{2}{r} A_{t}' &= 0,\label{eins00}\\ \nonumber \\
   U''+ \dfrac{2U'}{r}-\dfrac{6}{L^2}-\dfrac{\epsilon_{1}^{2}L^{2}}{2}\left(\dfrac{L^4 B^{2}}{r^4}+A_{t}'^{2}\right)&=0,\label{eins01}\\ \nonumber \\
    rU'+U-\dfrac{3 r^{2}}{L^{2}}+\dfrac{\epsilon_{1}^{2} L^{2}}{4}\left(\dfrac{L^{4}B^{2}}{r^{2}}+r^{2}A_{t}'^{2}\right) &= 0,\label{eins02}\\ \nonumber \\
         \partial_{x_{1}}^{2}A_{x_{2}} &= 0\label{eins04}.
\end{align}    
\end{subequations}
The equations (\ref{eins01}) and (\ref{eins02}) come from $R_{00}=R_{33}$ and $R_{11}=R_{22}$ components of Einstein equations respectively, whereas (\ref{eins00}) and(\ref{eins04}) arise from Maxwell equations. Considering (\ref{eins00}) and (\ref{eins01}), the solution of the coupled system is
\begin{equation}\label{pdeSol}
\begin{array}{rcl}
A_{t} &=& c_{2}-\dfrac{c_{1}}{r},\\ \\
U(r) &=& \dfrac{L^2 \epsilon_{1} ^2 \left(B^2 L^4+c_{1}^2\right)}{4 r^2}+\dfrac{r^2}{L^2}-\dfrac{c_3}{r}+c_4,
\end{array}
\end{equation}
being $c_{i}$ integration constants. Substituting the $U(r)$ expression into (\ref{eins02}) we obtain $c_{4}=0$ identically.
Using holographic prescriptions, we can associate $c_1$ with the charge density $Q$, and $c_2$ with the chemical potential $\mu$. Moreover, the remaining integration constant $c_{3}$, is associated to the mass of the black hole
\begin{equation}\label{c3}
c_{3}=\dfrac{\kappa_{4}^2}{\mathcal{V}_{2}} ML^{2} \equiv\dfrac{\mathcal{M}}{L^{2}},
\end{equation}
in which $\mathcal{V}_2$ stands for the volume of the spatial sector $(x_1,x_2)$,
obtained with the formalism founded in \cite{Balasubramanian:1999re} and detailed in the next section\footnote{As a consistency check, we computed the black hole mass with the quasi-local formalism building in \cite{Kim:2013zha} and obtained the same result.}. To analyze the roots of $U(r)$, let us write
\begin{equation}\label{Uinter}
\begin{array}{rcl}
U(r)&=&\dfrac{r^{2}}{L^{2}}-\dfrac{\mathcal{M}}{L^{2}}\dfrac{1}{r}+\dfrac{\mathcal{F}}{L^{2}}\dfrac{1}{r^{2}},\\ \\
\mathcal{B}&\equiv& B L^{4}/2, \ \ \ \ \mathcal{Q}\equiv L^{2}Q/2, \ \ \ \ \mathcal{F}\equiv \epsilon_{1}^{2}\left(\mathcal{B}^{2}+\mathcal{Q}^{2}\right).
\end{array}
\end{equation}
There are two real roots of this function denoting the inner and outer horizon of the black hole
\begin{equation}\label{roots}
r_{\pm}=\dfrac{\sqrt{\Omega}}{2}\left(1\pm \sqrt{\dfrac{2\mathcal{M}}{\Omega^{3/2}}-1}\right),
\end{equation}
where we have defined
\begin{equation}\label{definitions}
\Omega\equiv \dfrac{4\cdot 2^{2/3} \mathcal{F}+\delta^{2}}{2^{1/3}\cdot 3^{1/2}\delta},\ \ \ \ \delta\equiv\left(3^{3/2}\mathcal{M}^{2}+\sqrt{3^{3}\mathcal{M}^{4}-4^{4}\mathcal{F}^{3}}\right)^{1/3}.
\end{equation}
The extremal black hole is reached when the horizon positions merge into a single one, as can be seen from (\ref{roots}) 
\begin{equation}\label{omega2}
\Omega=2^{2/3}\mathcal{M}^{2/3}.
\end{equation}
On the other hand, the second term on r.h.s. of the $\delta$ expression implies
\begin{equation}\label{extremalC}
3^{3}\mathcal{M}^{4}\geq 4^{4}\mathcal{F}^{3}.
\end{equation}
A violation of this bound would result in a spacetime with a naked singularity\footnote{Curvature scalar $R=-\dfrac{r^2 U''(r)+4 r U'(r)+2 U(r)}{r^2}$.}, whereas its saturation implies
\begin{equation}
\delta=3^{1/2}\mathcal{M}^{2/3}.
\end{equation}
Substituting this limit value of $\delta$ in the $\Omega$ expression (\ref{definitions}) we obtain
\begin{equation}\label{extremal2}
3^{3}\mathcal{M}^{4}=4^{4}\mathcal{F}^{3};
\end{equation}
a relation between $\mathcal{M}$ and $\mathcal{F}$ that defines the extremal black hole \cite{zaanen_liu_sun_schalm_2015}.
The outer horizon in the non-extremal case is relevant regarding holographic setups\footnote{However, see \cite{Hartnoll:2020fhc} for an interesting incursion beyond the outer horizon, diving into the singularity.}. To identify it, we refer to Figure \ref{fig:outerh}.

Regarding the $t$-component of the vector potential, one must choose a gauge such that $A_{t}$ vanishes at the outer horizon, hence\footnote{This argument is taken from \cite{Gubser:2008px}: at the horizon, the one form $A_{t}=A_{t}(r)dt$ is ill-defined since has infinite norm there. As another equivalent statement, this boundary condition is imposed for the Wilson loop of the Maxwell field around the Euclidean thermal circle to be regular as $r \mapsto r_{+}$, where the circle shrinks to zero in the cigar geometry \cite{Hartnoll:2016apf}. } 
\begin{equation}\label{Qvsr}
\mathcal{Q}=\dfrac{L^2}{2}\mu r_{+}.
\end{equation}
Also, from $U(r_{+})=0$, the outer event horizon verifies
\begin{equation}
\mathcal{M} = r_{+}^{3}+\dfrac{\mathcal{F}}{r_{+}},
\end{equation}
which allows us to express the solution (\ref{pdeSol}) as (see \cite{Faulkner:2009wj,Romans:1991nq} for a similar parametrizations)
\begin{equation}\label{metricCoef}
\begin{array}{rcl}
 U(r) &=& \dfrac{r^{2}}{L^{2}}\chi(r),\\ \\
 \chi(r) &=& 1-\dfrac{r_{+}^{3}}{r^{3}}+\dfrac{\mathcal{F}}{r^{4}}\left(1-\dfrac{r}{r_{+}}\right),\\ \\
 A_{t}(r) &=& \mu\left(1-\dfrac{r_{+}}{r}\right),
\end{array}
\end{equation}
exhibiting the global AdS conformal factor $r^2/L^2$ and the \emph{blackening} factor $\chi (r)$.

The electric component of the Maxwell field acts as a source field that fixes the charge density at the boundary
\begin{equation}\label{rhoHolo}
\rho\equiv\langle J^{t}\rangle=\dfrac{1}{\mathcal{V}_{2}\beta_{T}}\left(\dfrac{\delta S_{on-shell}}{\delta A_{t}}\right)_{r\mapsto\infty}=\dfrac{\epsilon_{1}^{2}}{\kappa_{4}^{2}}\dfrac{\mu r_{+}}{2},\ \ \ \Rightarrow \ \ \rho=2\dfrac{\epsilon_{1}^{2}}{\kappa_{4}^{2}}\dfrac{\mathcal{Q}}{L^{2}},
\end{equation}
where $\beta_{T}$ is the inverse of Hawking temperature in the Euclidean periodic time. Regarding the Hawking temperature and extremality condition, we have
\begin{equation}\label{hawking}
    T=\dfrac{1}{4\pi}U'(r_{+})=\dfrac{3r_{+}^{4}-\mathcal{F}}{4\pi L^{2} r_{+}^{3}}.
\end{equation}
As we have already seen, the extremal black hole is reached when $r_{+}=\sqrt{\Omega}/2$ (see (\ref{roots}-\ref{definitions}). Using (\ref{omega2}) and (\ref{extremal2}), the latter equation gives exactly $3r_{+}^{4}=\mathcal{F}$, for which the Hawking temperature is $T=0$. It is self-evident that $\mathcal{F}$ diminishes the temperature relative to the neutral black hole for which the temperature scales linearly with the horizon position. Indeed, the limited behaviors
\begin{equation}\label{Tlimits}
  r_{+} \approx
    \begin{cases}
      \dfrac{4}{3}\pi L^{2}\, T & \text{if} \qquad  T \gg \left(\mu,\mathcal{B}\right),\\ \\
      \dfrac{1}{3}\pi L^{2}\, T+\dfrac{L^2}{2\sqrt{6}}\mu  & \text{if} \qquad T \ll \left(\mu,\mathcal{B}\right),\\ \\     
    \end{cases}       
\end{equation}
separate from the high-temperature states of the first line, described by the AdS-Schwarzchild black brane together with both, small electric and magnetic fields; and the second line, the low-temperature regime, where the horizon remains non-vanishing as the temperature goes to zero, resulting in a \emph{residual} entropy at zero-temperature\footnote{See, for instance \cite{Santos:2023mee}, in which the authors analyze the effects of this residual entropy in transport phenomena. Moreover, in \cite{Hartnoll:2016apf}, it was pointed out that SYK models can offer insights regarding the relation of residual entropy and degenerate states at zero temperature.}. In section \ref{machtIR} we will use this statement to capture relevant properties of the low energy superconductor in the near horizon geometry.
To end this section, a physical Hawking temperature demands that $3r_{+}^{4}\geq \mathcal{F}$ and represents a bound for allowed bulk values of $\mathcal{B},\mathcal{Q}$ and $T$ however, the dual quantities ($\mathcal{B}$, $\mu$, $T$) have different restrictions relative to the latter discussion. To prove the above statement, we need to get the proper thermodynamic potential and then, seek the thermodynamic stability condition of the system using the equation of state; as we will see shortly (the following section is based mostly on \cite{Hartnoll:2007ai}). 

\section{Thermodynamic stability}\label{stability}

In this section, we analyze the thermodynamic properties of the Dyonic solution (\ref{metricCoef}) working in the grand canonical ensemble, with chemical potential fixed at the boundary. The relevant potential $\mathcal{P} =- T\log\mathcal{Z}=-T S_{ren}^{os}$ is obtained from the renormalized Euclidean on-shell action\footnote{ 
We take
$$
S^{os}_{ren}=\dfrac{1}{\kappa_{4}^{2}}\int d^{4}x\sqrt{-g}\left(-\dfrac{6}{L^2}+\dfrac{L^2 \epsilon_{1}^{2}}{4}F^{2}\right)+\dfrac{1}{\kappa_{4}^{2}}\int d^{3}x\sqrt{-\gamma}\left( c_{1}\theta+\dfrac{c_{2}}{L}\right),
$$
being $\theta$ the extrinsic curvature $\theta=\gamma^{\mu\nu}\theta_{\mu\nu}$ with $\theta_{\mu\nu}=-\dfrac{1}{2}\left( \nabla_{\mu}n_{\nu}+\nabla_{\nu}n_{\mu}\right)$ and $n_{\mu}$ is an outward normal vector at $r=r_{0}$ slice with $\gamma$ the induced metric at that point. The constants $c_{i}$  are determined by the requirement to suppress the divergences when the surface defined by $r=r_{0}$  tends to infinity, see \cite{Balasubramanian:1999re}.
}
\begin{equation}
S_{ren}^{os}=\dfrac{\beta_{T}}{\kappa_{4}^{2}}\dfrac{\mathcal{V}_{2}}{L^4}\left(r_{+}^{3}+\dfrac{\epsilon_{1}^{2}}{ r_{+}}\left(\mathcal{Q}^{2}-3\mathcal{B}^{2}\right)\right).
\end{equation}
With the above equation, thermodynamic quantities are obtained straightforwardly, i.e., charge density and magnetization, whereas the entropy density is computed with the Bekenstein-Hawking area law
\begin{subequations}\label{termo}
   \begin{align}
    \rho &=-\dfrac{1}{\mathcal{V}_{2}}\dfrac{\partial \mathcal{P} }{\partial \mu}=\dfrac{1}{2}\dfrac{\epsilon_{1}^{2}}{\kappa_{4}^{2}}\mu r_{+}=2\dfrac{\epsilon_{1}^{2}}{\kappa_{4}^{2}}\dfrac{\mathcal{Q}}{L^{2}},\label{rho}\\ \nonumber \\
   \tilde{m} &= -\dfrac{1}{\mathcal{V}_{2}}\dfrac{\partial \mathcal{P}}{\partial B}=-\dfrac{3}{2}\dfrac{\epsilon_{1}^{2}}{\kappa_{4}^{2}}\dfrac{L^4}{r_{+}}B=-3\dfrac{\epsilon_{1}^{2}}{\kappa_{4}^{2}}\dfrac{\mathcal{B}}{r_{+}},\label{mag}\\ \nonumber \\
    s &= \dfrac{4\pi}{\kappa_{4}^{2}}\dfrac{r_{+}^{2}}{L^{2}}.\label{entropy}       
\end{align}    
\end{subequations}
The equation (\ref{rho}) obtained with thermodynamical arguments, is equivalent to (\ref{rhoHolo}), being the later computed according to the holographic precepts.  

Associated with the time translation symmetry, there is a conserved charge\footnote{We will follow the procedure outlined in \cite{Hartnoll:2007ai} using the stress tensor as a functional of the boundary metric. See also \cite{Balasubramanian:1999re} for a more complete theoretical treatment.}
\begin{equation}\label{noether}
\mathsf{Q}_{\zeta}=\int d^{2}x\sqrt{\sigma}\zeta^{\mu}T_{\mu\nu}k^{\nu},
\end{equation}
where the integration is performed over the spatial boundary directions with metric $\sigma_{ij}$ and $\zeta^{\mu}$, $k^{\nu}$ correspond to the time-like vector orthogonal to the surface $t$-constant and the Killing vector for time translations, respectively. Energy-momentum tensor at the AdS boundary reads
\begin{equation}
\langle T_{\mu\nu}\rangle=\dfrac{2}{\kappa_{4}^{2}}\left(\theta_{\mu\nu}-\theta\gamma_{\mu\nu}-\dfrac{2}{L}\gamma_{\mu\nu}\right).
\end{equation}
We interpret the charge (\ref{noether}) associated with time translation as the mass $M$ of the black hole. Hence, the constant $c_{3}$ in (\ref{pdeSol}) turns out to be (\ref{c3}) with energy density
\begin{equation}
\dfrac{M}{\mathcal{V}_{2}}\equiv\varepsilon = \dfrac{2}{\kappa_{4}^{2}L^4}\left(r_{+}^{3}+\dfrac{\mathcal{F}}{r_{+} }\right).
\end{equation}
Local thermodynamic stability of the background (\ref{theMetric}) can be analyzed if we consider the equation of state $\varepsilon (\rho,s)$ for the grand canonical ensemble with $\mathcal{B}$ (or $B$)  as an external parameter. Indeed, using (\ref{hawking}), (\ref{rho}) and (\ref{entropy}), energy density takes the form
\begin{equation}
\varepsilon  (\rho,s)= \dfrac{\kappa_{4}}{4L}\left(\dfrac{s}{\pi}\right)^{3/2}\left[1+\dfrac{16\pi^{2}\epsilon_{1}^{2}}{L^{4}\kappa_{4}^{4}}\left(\dfrac{\mathcal{B}}{s}\right)^{2}+\dfrac{4\pi^{2}}{\epsilon_{1}^{2}}\left(\dfrac{\rho}{s}\right)^{2}\right],
\end{equation}
The condition over the Hessian: $\det\left[\partial^{2}_{\rho,s}\varepsilon(\rho,s)\right]>0$, implies positive energies which can be accomplished if
\begin{equation}\label{bmuvalues}
3 r_{+}^4+\left(3\mathcal{B}^{2}+\mathcal{Q}^{2}\right)\epsilon_{1}^{2}>0,
\end{equation}
which is certainly true. Hence, the dual system is thermodynamically stable for all values $(\mathcal{B},\mu, T)$. In section \ref{machtIR} we will compare the effect of the magnetic field on the Thermodynamics characterized by the grand canonical ensemble with variables $T$ and $\mu=2\mathcal{Q}/L^{2} r_{+}$, focusing on the limit $ T\mapsto 0$, for which AdS$_{2}$ governs the dynamics \citep{Faulkner:2009wj}. To end this section, the first law of Thermodynamics 
\begin{equation}
d\varepsilon = Tds-\mu d \rho,
\end{equation}
is satisfied using the set of formulae above\footnote{As argued in \cite{Hartnoll:2007ai}, it might be thought that the term $B\tilde{m}$ is required in the expression of the thermodynamic potential. However, it is consistent because we treat the $B$ field as an external parameter and without this term, the Gibbs potential is satisfied.}.
\section{Instability and AdS$_{2}$ throat}\label{throat}
Recall that the hair formation is related to spontaneous symmetry breaking and phase transition from a vanishing scalar field solution to a non-trivial profile \cite{Gubser:2008px}. In addition, in the AdS$_{2}$ throat, the condensate is stable due to a strong gravitational attraction, as shown in \cite{Hartnoll:2008kx}. A lesson we have learned from these works is that the near horizon geometry is sufficient for hair formation, even if the scalar field is neutral. We expect the same behavior in our dyonic black hole. In this sense, close to extremality, where the relation \eqref{extremal2} holds, our background develops an AdS$_{2}$ geometry near the horizon, hence, the perturbative scalar field condenses in this regime. From (\ref{hawking}), the extremal limit reduces the metric function to
\begin{equation}\label{extremalU}
U_{ext}(r)=\dfrac{r^2}{L^2}+\dfrac{3 r_{+}^4}{L^2 r^2}-\dfrac{4 r_{+}^3}{L^2 r}.
\end{equation}
Expanding near $r_{+}$, the function $U_{ext}$ develops a double zero due to the two nonzero real roots that define extremality. Indeed, under these circumstances, the metric (\ref{theMetric}) asymptotes to
\begin{equation}\label{NHmetric}
ds^{2}\approx -\dfrac{6}{L^2}u^{2}dt^{2}+\dfrac{L^{2}}{6}\dfrac{du^{2}}{u^{2}}+\dfrac{r_{+}^{2}}{L^2}d\vec{x}^{2},
\end{equation}
where $u=r-r_{+}$. An AdS$_{2}\times\mathbb{R}^{2}$ arise with effective curvature $L_{eff}=L/\sqrt{6}<L$. On the other hand, consider the perturbative scalar field equation (\ref{eins3})
\begin{equation} \label{masterscalar}
   \dfrac{1}{\sqrt{-g}}\partial_{r}\left[\sqrt{-g}g^{rr}\partial_{r}\psi\right]+g^{\tilde{i}\tilde{i}}\Delta\psi
   - 2\textit{i} A_{x_{2}}g^{\tilde{i}\tilde{i}}\partial_{x_{2}}\psi
    -\left[g^{tt}A_{t}^{2}+g^{\tilde{i}\tilde{i}}A_{x_{2}}^{2}\right]\psi- m^{2}\psi=0,  
\end{equation}
in which $\Delta=\partial_{1}^{2}+\partial_{2}^{2}$ and $g^{\tilde{i}\tilde{i}}\equiv g^{11}=g^{22}$. In addition, the London gauge $A=Bx_{1}dx_{2}$ and $A_{t}$ come from the background fields. This equation gives rise to a minimal formation of an Abrikosov vortex structure provided we assume $\psi=\psi(r,x_{1},x_{2})$.   Hence, (\ref{masterscalar}) reduces to
\begin{equation}\label{scalarmastereq1}
    \partial_{r}\left[\dfrac{r^{2}}{L^{2}}U(r)\partial_{r}\psi\right]+\dfrac{r^{2}A_{t}^{2}}{L^{2}U(r)}\psi-\dfrac{m^2r^2}{L^2}\psi = 
    -\Delta\psi+2\textit{i}Bx_{1}\partial_{x_2}\psi+B^{2} x_{1}^{2}\psi.
\end{equation}
It is licit to assume the functional dependency $\psi=e^{inx_{2}}\gamma(x_{1})R(r)$ under which the above equation is separable. Let $\lambda$ be the separation parameter, then (\ref{scalarmastereq1}) yields the following system of equations
\begin{subequations}\label{separation}
	\begin{align}
        \partial_{r}\left(\dfrac{r^{2}}{L^{2}}U\partial_{r}R\right)+\dfrac{1}{U}\left(\dfrac{r A_{t}}{L}\right)^{2}R-\left(\dfrac{mr}{L}\right)^{2}R &=B \lambda R \label{Req1}, \\ \nonumber\\
        -\left( \partial_{1}^{2}-p^{2}\right)\gamma -2Bpx_{1}\gamma+\left(Bx_{1}\right)^{2}\gamma &= B \lambda\gamma.\label{gammaeq1}
	\end{align}
\end{subequations}
The vortex structure is revealed by (\ref{gammaeq1}) and, according to holographic renormalization, (\ref{Req1}) gives us the scale on boundary directions (depending on the value of holographic coordinate $r$) at which the dynamics take place.

To show the instability close to the horizon through AdS$_{2}$ throat, we use our dyonic solution (\ref{metricCoef}). At the extremal limit, which occurs at $\mathcal{F}=3r_{+}^{4}$, the near horizon metric is captured by (\ref{NHmetric}). Plugging the latter into the radial part of the scalar equation (\ref{Req1}) and considering $A_{t}$ in a series expansion, we get\footnote{In \cite{Horowitz:2009ij} the background has only electric charge and the scalar condensate is purely real.}
\begin{equation}\label{effectiveR}
R_{eff}''(u)+\dfrac{2}{u}R_{eff}'(u)-\dfrac{L_{eff}^{2}m_{eff}^{2}}{u^{2}}R(u)=0
\end{equation}
where $u=r-r_{+}$ and
\begin{equation}\label{effectivemass}
m_{eff}^{2}\equiv m^{2}-\dfrac{6L_{eff}^{2}}{r_{+}^{2}}\left(\dfrac{\mu^{2}}{6}-\lambda B\right).
\end{equation}
The instability of the bulk to form scalar hair in the near horizon region is accomplished if the effective mass verifies
\begin{equation}\label{BFbounds}
m_{eff}^{2} L_{eff}^{2} \leq \left(-\dfrac{1}{4}\right)_{\text{AdS}_{2}},
\end{equation}
i.e., below the Breitenlohner-Freedman (BF) bound in the AdS$_{2}$ throat\footnote{Similar bound of effective AdS$_{2}$ equation is widely used in the theory of holographic quantum critical points, e.g., \cite{Faulkner:2009wj}.}. The bare mass must be above the BF bound of the four-dimensional bulk
\begin{equation}
L^{2}m^{2} \geq \left(-\dfrac{9}{4}\right)_{\text{AdS}_{4}}.  
\end{equation}
In the next section, we choose the value $m^2L^2=-2$ at the UV boundary (implying both dual operators normalizable). Aside, in section \ref{final}, we will explore specific values for the effective mass (\ref{effectivemass}) in the AdS$_2\times\mathbb{R}^{2}$ black hole.
\section{Superconductivity with scalar perturbation}\label{vortexsection}

Consider the scalar field action 
\begin{equation}\label{scalarAction}
S_{\psi}=-\dfrac{\epsilon_{2}^{2}L^{2}}{\kappa_{4}^{2}}\int d^{4}x\sqrt{-g}\left(\vert D\psi\vert^{2}+m^{2}\vert\psi\vert^{2}\right),
\end{equation}
that has been effectively decoupled from the Einstein-Maxwell sector (see section \ref{decouplinS}). The program of Abelian-Higgs holographic superconductivity rests on the Hawking temperature-dependency of the dual operators at the boundary
\begin{equation}\label{dualOPS}
\langle\hat{\mathcal{O}}_{1}\rangle = \dfrac{\epsilon_{2}^{2}}{\kappa_{4}^{2}}\dfrac{a(x)}{L^{2}}, \ \ \ \ \langle\hat{\mathcal{O}}_{2}\rangle = \dfrac{\epsilon_{2}^{2}}{\kappa_{4}^{2}}\dfrac{b(x)}{L^{2}},
\end{equation}
obtained by renormalizing and evaluating the on-shell action (\ref{scalarAction}). In the above expressions, $a(x)$, $b(x)$ are the coefficients of the full scalar field equation expanded at the boundary, which depends on the $(x_{1},x_{2})$-coordinates in our setup. One of the main problems that arise in the context of charged black holes, which is our case, is that the function $r_{+}=r_{+}( T,\mathcal{Q},\mathcal{B};L)$ is highly non-trivial (see \ref{hawking}). However, as we shall see in section (\ref{matchinProcedure}), with the \emph{matching} method we can obtain a slightly more amenable expression if the horizon position $r_{+}$ is considered as one of the limiting cases (\ref{Tlimits}). With this procedure, we arrive at phase space expressions of the form $f\left(\mathcal{B},\mathcal{Q}, T; r_{m}\right)=0$ being $r_{m}$ some point of the bulk between the boundary and the horizon. Recall that in our decoupling limit, we can not capture the thermal behavior of the dual VEV order parameter. Therefore, the analysis of the superconductor is limited in two situations: near the critical temperature $T_{c}\propto\sqrt{\mathcal{Q}}$ for which the order parameter tends to zero \cite{Hartnoll:2008kx}, and the other limit, when the temperature tends to zero. The last case is studied in section \ref{machtIR}. Although we were not able to find a phase space, we found sufficient evidence that reinforces the existence of an order parameter in that limit. 

\subsection{Landau Levels}

A vortex lattice structure can be built in our gravitational configuration, which will be dual to an Abrikosov superconductor. First, by considering the equation (\ref{gammaeq1}) and defining $y\equiv \sqrt{B}\left(x_{1}-p/B\right)$, being $p$ an arbitrary constant\footnote{From previous experience, we know that the construction of the Abrikosov lattice requires $p$ to be quantized \cite{ Montull:2009fe,Maeda:2009vf}.}, we get
\begin{equation}\label{gammafunction}
-\dfrac{d^{2}\gamma}{d y^{2}}+y^{2}\gamma=\lambda\gamma.
\end{equation}
Identifying the above equation with the harmonic oscillator one that emanates from the linearized Ginzburg-Landau equation, the solutions are given in terms of Hermite polynomials, namely
\begin{equation}\label{hermite}
\gamma_{n}(y;p)=e^{-y_{p}^{2}/4}H_{n}(y),
\end{equation}
with eigenvalues $\lambda_{n}=2n+1$. The value $n=0$ corresponds to the \emph{lowest Landau level} (LLL). Therefore, the functions in (\ref{dualOPS}) are proportional to $\exp\left(\frac{ipx_{2}-y_{p}^{2}}{4}\right) H_{n}(y)$ for the operator $\mathcal{O}_{2}$. We restrict ourselves to the standard quantization of $\mathcal{O}_{2}$ case for the rest of this report nevertheless, both operators in (\ref{dualOPS}) are normalizable, allowing an alternative quantization too \cite{zaanen_liu_sun_schalm_2015}.
 
\subsection{Phase space of superconducting state}\label{matchinProcedure}

The solution for the $R(r)$ equation in (\ref{Req1}) has information on the renormalization group flow (through $r$-coordinate) of the superconducting system, from the boundary UV to the IR degrees of freedom. We use the matching method procedure that links at some bulk point, the boundary, and the near horizon expansions of the $R(r)$ equation. A phase space $\left( \mathcal{B}, T\right)$ parametrized by the electric charge $\mathcal{Q}$ is driven from it\footnote{The matching method has been implemented in a series of papers that analyze the holographic superconductor within the Abelian-Higgs theory. See \cite{Gregory:2009fj} for a representative example.}. Starting with (\ref{Req1}) and performing the change of coordinate $z=r_{+}/r$ to have better control of the location of the horizon and boundary system, the metric function $U(r)$ in (\ref{theMetric}) and the electric component $A_{t}$ read
\begin{equation}\label{Metric3}
\begin{array}{rcl}
U(z) &=&  \dfrac{r_{+}^{2}}{L^2 z^2} \chi(z), \\ \\
\chi(z) &=& 1-z^{3}+\dfrac{\mathcal{F}}{r_{+}^{4}}\left(z-1\right)z^{3}, \\ \\
A_{t}(z) &=&\mu\left(1-z\right).
\end{array}
\end{equation}
In the language of this coordinate, the boundary is located at $z=0$ whereas $z=1$ corresponds to the outer horizon $r_{+}$. Besides, taking $R(r)\mapsto R(z)$ the equation (\ref{Req1}) is now
\begin{equation}\label{Zeq}
R''(z)+\left(\dfrac{\chi'(z)}{\chi(z)}-\dfrac{2}{z}\right)R'(z)+\dfrac{1}{\chi(z)}\left(\dfrac{ L^4}{r_{+}^{2}\chi(z)} A_{t}(z)^{2}-\dfrac{m^2 L^2}{z^{2}}-\dfrac{L^{4}\lambda_{n}\mathcal{B}}{r_{+}^{2}}\right)R(z)=0.
\end{equation}
Near the boundary, (\ref{Zeq}) behaves as
\begin{equation}\label{bdyvalues}
R(z\mapsto 0)=\alpha z + \beta z^2
\end{equation}
from which we can read off the conformal dimension $\Delta=1$ and $\Delta=2$ for $\mathcal{
O}_{1}\sim \alpha$ and $\mathcal{O}_{2}\sim \beta$ respectively; both of them associated with $m^{2}L^{2}=-2$. At the other limit, near the horizon, we perform a series expansion around $z=1$
\begin{equation}\label{seriesU}
R(z)=R(1)+R'(1)\left(z-1\right)+\dfrac{1}{2}R''(1)\left(z-1\right)^{2}+...\ .
\end{equation}
Regularity at the horizon implies $R(1)<\infty$ and, up to the second order, the coefficients turn out to be\footnote{A second integration constant in this near horizon limit weights a logarithmic term, which is divergent at $z=1$ hence, we drop it.}
\begin{align}\label{Zexp}
\dfrac{R(z)}{R(1)} &=1+\dfrac{2r_{+}^{2}\left(r_{+}^{2}-\lambda_{n}\mathcal{B}\right)}{3r_{+}^{4}-\mathcal{F}}\left(z-1\right)\\ \nonumber
&- \dfrac{ r_{+}^4 \left(\mathcal{Q}^2-\lambda_{n} ^2 \mathcal{B}^2-3 \mathcal{F}\right)+2 r_{+}^6 \lambda_{n}  \mathcal{B}+2 r_{+}^2 \lambda_{n}  \mathcal{B} \mathcal{F}+2 r_{+}^8  }{\left(3r_{+}^{4}-\mathcal{F}\right)^{2} }\left(z-1\right)^{2}+... .
\end{align}
A second-order expansion near the horizon will be enough to observe the onset of the superconducting state in the canonical ensemble, as we shall see next.

\subsection{Fixed charge density}

In Abelian-Higgs holography, the critical temperature is proportional to (square root of) the charge density $\rho$. In this work, we present such temperature in terms of the black hole electric charge $\sqrt{\mathcal{Q}}$, exploiting the fact that are proportional to each other, as we can see from (\ref{rho}). To achieve the matching in the canonical ensemble, we need to assume that $\beta$ in (\ref{bdyvalues}) is small due to proximity from below to the critical temperature $T_{c}$; a condition that is observed in experimental setups of type-II superconductors. Hence, the strategy is to take the approximation in (\ref{Zexp}) and consider the high-temperature limit $ T\gg (\mathcal{Q},\mathcal{B})$, then, match it with the boundary values (\ref{bdyvalues}) choosing the source $\alpha=0$. We obtain, for $\mathcal{O}_{2}$ case
\begin{equation}\label{matchCanonical}
\begin{array}{rcl}
& &\left(4 r_{+}^6 \lambda_{n}  \mathcal{B}+r_{+}^4 (\lambda_{n}  \mathcal{B}-\mathcal{Q}) (\mathcal{Q}+\lambda_{n}  \mathcal{B})-4 r_{+}^2 \lambda_{n}  \mathcal{B} \mathcal{F}-r_{+}^4 \mathcal{F}+r_{+}^8+\mathcal{F}^2\right)\\ \\
&+& \left(6 r_{+}^2 \lambda_{n}  \mathcal{B} \left(\mathcal{Q}^2+\mathcal{B}^2\right)-8 r_{+}^4 \left(\mathcal{Q}^2+\mathcal{B}^2\right)-2 r_{+}^6 \lambda_{n}  \mathcal{B}+2 r_{+}^4 (\mathcal{Q}+\lambda_{n}  \mathcal{B}) (\mathcal{Q}-\lambda_{n}  \mathcal{B})+10 r_{+}^8\right)z_{m}\\ \\
&+&\left( -2 r_{+}^2 \lambda_{n}  \mathcal{B} \left(\mathcal{Q}^2+\mathcal{B}^2\right)+3 r_{+}^4 \left(\mathcal{Q}^2+\mathcal{B}^2\right)-2 r_{+}^6 \lambda_{n}  \mathcal{B}-r_{+}^4 (\mathcal{Q}-\lambda_{n}  \mathcal{B}) (\mathcal{Q}+\lambda_{n}  \mathcal{B})-2 r_{+}^8 \right)z_{m}^{2}\\ \\
&=&\left(\mathcal{F}-3 r_{+}^{2}\right)^{2}\dfrac{\beta z_{m}^{2}}{R(1)},
\end{array}
\end{equation}
where we have to use the definition of $\mathcal{F}$ in (\ref{Uinter}) taking $\epsilon_{1}=1$. In addition, to ensure the continuity of the matching in the entire bulk, we must take into account the matching of the first derivative of $R(z)$. With these considerations,
we obtain an algebraic system for $\lbrace\mathcal{B},\beta\rbrace$
. Furthermore, taking the LLL, for which $\lambda_{0}=1$, the upper critical magnetic field as a function of temperature turns out to be
\begin{align}\label{eq:Bc2}
 \mathcal{B}_{c2}&= \dfrac{\mathcal{Q}}{2\sqrt{(1-z_m)(5z_m+1)}}\left(\dfrac{T}{T_c}\right)^2\Biggl[z_m-4\\ 
      &\quad+ \sqrt{ 3(4+z_m (7 z_m-8))+4 (1-z_m) (5 z_m+1)\left(\dfrac{T_c}{T}\right)^4 }\Biggr],\nonumber \\ 
      T_c &=\dfrac{3}{{4 \pi L^2}} \left(\dfrac{1-z_m}{5 z_m+1}\right)\sqrt{\mathcal{Q}}\nonumber.
\end{align}

\begin{figure}[h]
    \centering
    \includegraphics[width=0.6\textwidth]{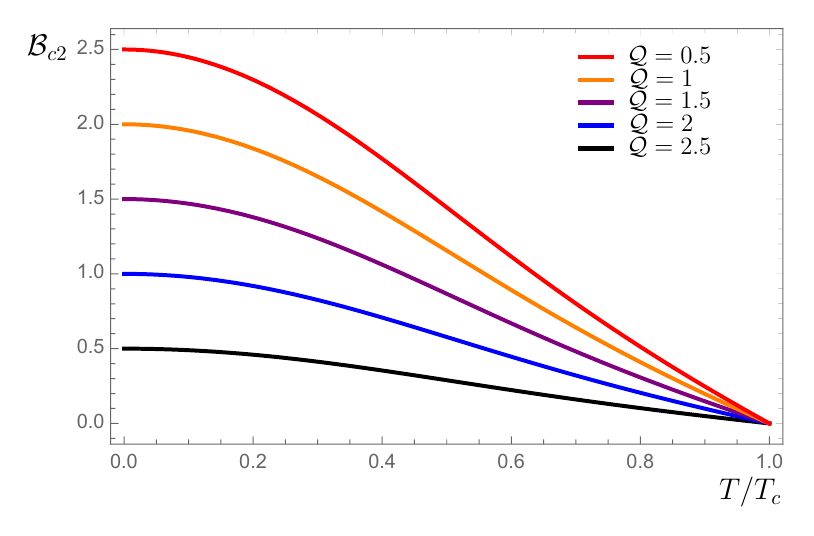}
    \caption{ The critical magnetic field $\mathcal{B}_{2}$ as a function of the temperature in the canonical ensemble where $\mathcal{Q}\propto\rho$ is considered fixed at the boundary, showing the regions in which the superconductivity is present; the lower left region, representing the onset of superconductivity in the canonical ensemble. Above these curves, there is no superconductivity. These curves should be compared with the ones obtained in \cite{Hartnoll:2008kx}. In this plot we take $z_{m}=0.9$ and $A_{t}=\left(1-z\right)\rho/r_{+}$, i.e., we consider $\kappa_{4}^{2}/\epsilon_{1}^{2}=1$.}
    \label{fig:BvsTcanonical}
\end{figure}

In Figure \ref{fig:BvsTcanonical}, we illustrate the relationship between the upper critical magnetic field $\mathcal{B}_{c2}$ as a function of temperature. We observe that a linear behavior emerges as the temperature approaches $T_{c}$, such a behavior is aligned with both the Ginzburg-Landau theory for type-II superconductors and experimental observations. We invite the reader to confront the behavior depicted in Figure 2 with those found in Abelian-Higgs holographic theory, see \cite{Zhao:2013pva}, for instance\footnote{In \cite{Zhao:2013pva} the dual theory is a nonrelativistic one with a Lifshitz scaling exponent.}.
\subsection{Fixed chemical potential}
In the grand canonical ensemble, where the chemical potential $\mu$ is fixed at the boundary, the equation \eqref{Qvsr} introduces additional complexity compared to the canonical ensemble, where $\mathcal{Q}$ is considered fixed. This complexity becomes evident in equation (\ref{matchCanonical}), where, for the grand canonical scenario, more substitutions involving $\mu$ are necessary for $\mathcal{F}$ and $r_{+}$. Despite this technical challenge, we managed to derive an expression for the phase space $f\left( \mathcal{B},\mu,T; z_{m}\right)=0$.

\begin{figure}[ht]
	\centering
		\includegraphics[width=.6\textwidth]{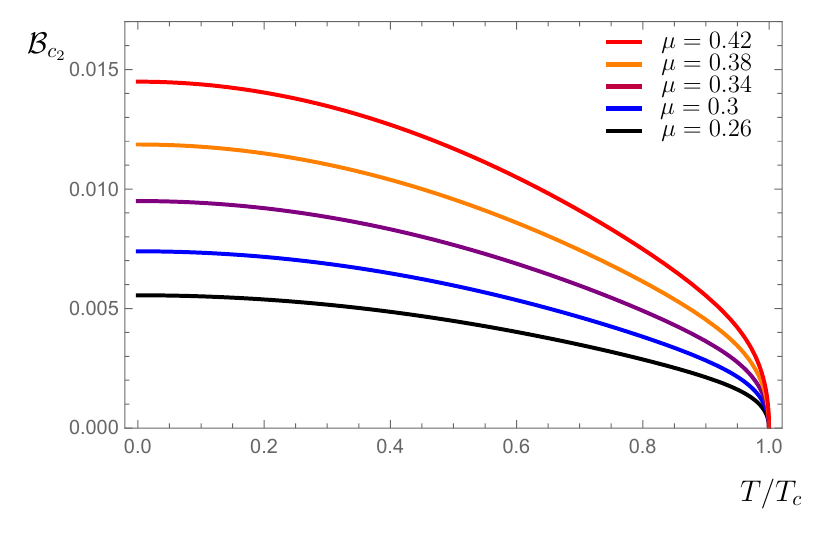}    
		\caption{The behavior of the upper critical magnetic field for the operator $\mathcal{O}_{2}$ as a function of temperature and different values of the chemical potential in the high-temperature limit $T \gg \left(\mu,\mathcal{B}\right)$ regime. Here we take $L=1$ and $z_{m}=0.9$. It can be appreciated qualitative changes in the critical magnetic field close to $ T=T_{c}$ relative to the canonical ensemble.}
		\label{fig:Bc2b}
    \label{fig:firstOrder_TvsB}
\end{figure}

Therefore, considering the second-order expansion of $R(z)$ equation and again, regarding the operator $\mathcal{O}_{2}$, the matching procedure yields
\begin{equation}\label{zm_op2}
\begin{array}{rcl}
\beta z_{m}^{2} &=& 1-\dfrac{r_{+}^{2}\left(L^{4}\lambda_{n}\mathcal{B}-2r_{+}^{2}\right)}{3r_{+}^{2}-\mathcal{F}}\left(z_{m}-1\right)\\ \\
&+& \dfrac{r_{+}^{2}}{4}\dfrac{\left(2r_{+}^{2}-\lambda_{n}\mathcal{B}L^{4}\right)\left(5r_{+}^{4}+\mathcal{F}-r_{+}^{2}L^{4}\lambda_{n}\mathcal{B}\right)}{\left(3r_{+}^{4}-\mathcal{F}\right)^{2}}\left(z_{m}-1\right)^{2}\\ \\
&-&\dfrac{r_{+}^{2}}{4}\dfrac{\left(18r_{+}^{6}-10r_{+}^{2}\mathcal{F}-3r_{+}^{4}L^{4}\lambda_{n}\mathcal{B}+3\mathcal{F}L^{4}\lambda_{n}\mathcal{B}+r_{+}^{4}L^{4}\mu^{2}\right)}{\left(3r_{+}^{4}-\mathcal{F}\right)^{2}}\left(z_{m}-1\right)^{2}.
\end{array}
\end{equation}
representing a phase space $f\left( \mathcal{B},\mu;z_{m},r_{+}\right)=0$ on the entire space-time. As in the canonical ensemble case, we use the high-temperature limit in (\ref{Tlimits}) and follow the same procedure, yielding
\begin{align}\label{secondO}
         \mathcal{B}_{c2} &=  \dfrac{\mu^2 L^4}{24\left(3\left(z_m-2\right)+\left(5z_m-2\right)\dfrac{T^2}{T_{c}^2}\right)}\Biggl[3\left(12-9z_m+\left(z_m-4\right)\dfrac{T^2}{T_{c}^2}\right)\left(1-\dfrac{T^2}{T_{c}^2}\right)\nonumber\\ 
         &-\sqrt{3}\Biggl[\Biggl(-27\left(24+z_m\left(9z_m-28\right)\right)+9\left(88+z_m\left(21z_m-68\right)\right)\dfrac{T^2}{T_{c}^2}\nonumber\\
         &-3\left(104+z_m\left(59z_m-52\right)\right)\dfrac{T^4}{T_{c}^4}\nonumber\\
         &+\left(40+z_m\left(103z_m-44\right)\right)\dfrac{T^6}{T_{c}^6}\Biggr) \Biggl(-1+\dfrac{T^2}{T_{c}^2}\Biggr)\Biggr]^{1/2}\Biggl],\\
         T_{c} &=\frac{\sqrt{3}  }{8 \pi }\mu. \nonumber
\end{align}
For this scheme, Figure \ref{fig:Bc2b} depicts the behavior of the upper critical magnetic field as a function of temperature. Here we observe important changes compared with the previous description in the canonical ensemble. In particular, the most interesting difference is the nonlinear behavior of the curves near the critical temperature $T_{c}$.
To the best of the authors' knowledge, the behavior of the upper critical magnetic field in the grand canonical ensemble, as depicted in Figure \ref{fig:Bc2b}, is typically not addressed in holographic superconductivity reports. We suspect that this omission is primarily due to technical difficulties rather than any other aspect.

\section{Diving into AdS$_{2}$}\label{machtIR}

As discussed in Section \ref{throat}, the near extremal limit of the dyonic solution (\ref{metricCoef}) exhibits an AdS$_2\times\mathbb{R}^{2}$ topology near the horizon. This aspect underpins the exploration of criticality and quantum phase transitions within holography \cite{Faulkner:2009wj, Faulkner:2010gj}. In this section, our focus shifts towards the near horizon and near extremal description of our perturbative scalar field model. To accomplish this, we analyze the Schrödinger potential corresponding to the radial part of the scalar field equation (\ref{Zeq}) in search of bound states distributed along the holographic coordinate. We interpret their presence as indicative of the density of Cooper pairs in the dual system. The significance of potential minima as a cutoff, separating the UV and IR degrees of freedom, is deferred to future investigations\footnote{It was pointed out in \cite{Faulkner:2010jy} that, integrating out the UV region of space-time to an IR cutoff, a Wilsonian renormalization group flow interpretation can be considered, extracting the low energy behavior. This interpretation is justified by the standard holographic UV/IR connection dictionary. Furthermore, in the AdS$_{2}$ region, a matching method is applied in a seminal work, to extract the IR behavior of the retarded Green functions at low frequency for some operator $\mathcal{O}$, dual to bulk fermion (and scalar) field \cite{Faulkner:2009wj}, with the role of renormalization group fixed point as well. See also \cite{Faulkner:2010gj} in which the authors add multitrace deformations in the boundary action and prove new kinds of instabilities. }. For a review of these topics, we invite the reader to consult, for instance, \cite{zaanen_liu_sun_schalm_2015,Hartnoll:2016apf}.


\subsection{Schrödinger potential}\label{cutoffsec}
For a scalar field mass satisfying the bound (\ref{BFbounds}), the equation (\ref{Zeq}) can be set into a Schrödinger-like equation if we consider $R(z)\mapsto p(z)\mathcal{Z}(z)$, being the integrating factor 
\begin{equation}\label{intFac}
p(z)=\dfrac{z}{\sqrt{\chi(z)}}.
\end{equation}
Then, the equation \eqref{Zeq} can be transformed into a Schr\"{o}dinger equation for the function $\mathcal{Z}$, which reads
\begin{equation}
\mathcal{Z}''-V(z)\mathcal{Z}=0,
\end{equation}
with a Schrödinger potential 
\begin{equation}{\label{potentialJi}}
 V\left(z\right)=\dfrac{1}{2}\dfrac{\chi''}{\chi}-\dfrac{1}{4}\left(\dfrac{\chi'}{\chi}\right)^{2}-\dfrac{\chi'}{z\chi}+\dfrac{2}{z^{2}}-\dfrac{1}{\chi}\left(\dfrac{L^{4}A_{t}^{2}}{r_{+}^{2}\chi}-\dfrac{2\mathcal{B}}{r_{+}^{2}}-\dfrac{m^{2}L^{2}}{z^{2}}\right).
\end{equation}
After replacing the form of the metric function and the gauge field, we can easily identify the singularities contained in this potential. Let us write
\begin{equation}\label{potentialz}
    V(z)=-\dfrac{a_0+a_2z^2+a_3 z^3+ a_4 z^4+a_5 z^5+a_6 z^6}{4 r_{+}^{6}z^{2}\left(z-1\right)^{2}\left(1+z+z^2-z^3 F\right)^{2}},
\end{equation}
where $F\equiv \mathcal{F}/r_{+}^{4}$ and the $a_{i}$-coefficients defined as
\begin{equation}
\begin{array}{rcl}
    a_0 &=& -4r_{+}^{6}\left(L^{2}m^{2}+2\right), \\[2pt]
    a_2 &=& 8\left(2\mathcal{Q}^{2}-r_{+}^{4}\mathcal{B}\right), \\[2pt]
    a_3 &=& 4\left(r_{+}^{6}\left( \left(F+1\right)\left(m^{2}L^{2}+4\right)-8\mathcal{Q}^{2}\right)\right), \\[2pt]
    a_4 &=& -4\left(r_{+}^{6}\left(m^{2}L^{2}+6\right)-4\mathcal{Q}^{2}\right),\\[2pt]
    a_5 &=& 8 r_{+}^4 (F+1) \mathcal{B}, \\[2pt]
    a_6 &=& r_{+}^{4}\left(r_{+}^{2}\left(1+2F+F^{2}\right)-8F\mathcal{B}\right).
\end{array}    
\end{equation}
In the form of (\ref{potentialz}), we can recognize two singularities appearing at the horizon $z=1$ and the boundary $z=0$. However, when taking the particular value of the mass $L^{2}m^{2}=-2$, the coefficient $a_{0}$ vanishes, and the singularity at the boundary is no more present. Choosing this value allowed us to identify the conformal dimension $\Delta_{1,2}=1,2$ in the boundary expansion of (\ref{Zeq}) (see (\ref{bdyvalues})).

\begin{figure}[tbp]
	\centering
	\begin{subfigure}[b]{.49\textwidth}
		\centering
		\includegraphics[width=.95\textwidth]{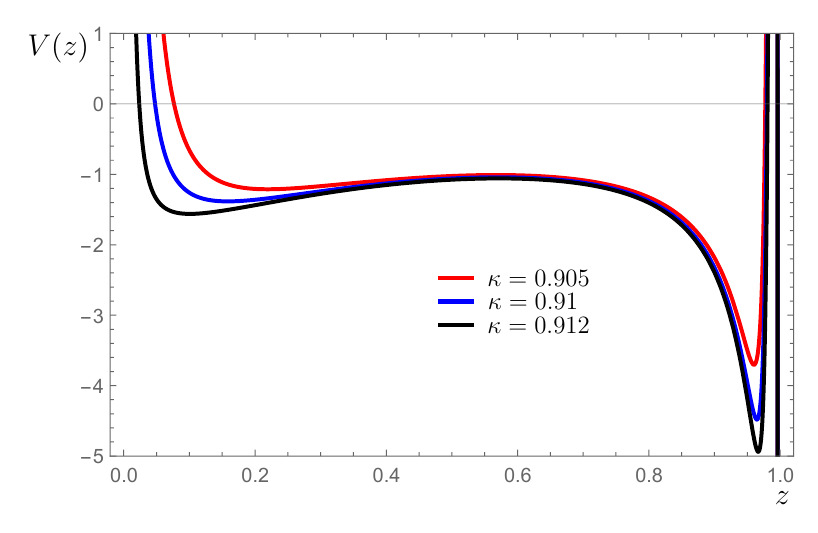}
		\caption{}
		\label{fig:Pot1a}
	\end{subfigure}
	\hfill     
	\begin{subfigure}[b]{.49\textwidth}
		\centering
		\includegraphics[width=.95\textwidth]{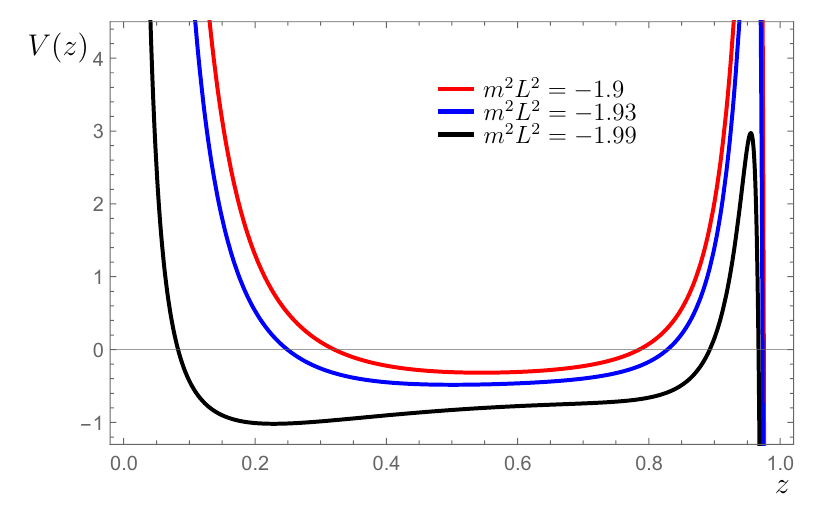}    
		\caption{}  
		\label{fig:Pot1b}
	\end{subfigure}
	\caption{The potential function (\ref{potentialJi}) for different values of the scalar field mass. (a): In this case we take $\mathcal{Q}\approx 0.62$ and $\mathcal{B}\approx 0.7$, close to extremality $\nu=2.99$. The minimum of the potential occurs at $z_{c}\approx 0.95$ for the three cases and represents the IR superconductor. (b): The potential function for $\nu=2.9$ and $\kappa\approx 0.91$. As we see, the potential is highly sensitive to the parameters as we move away from the extremality. However, bound states exist away from the horizon region.}
    \label{fig:pot1}
\end{figure}

\begin{figure}[tbp]
	\centering
	\begin{subfigure}[b]{.49\textwidth}
		\centering
		\includegraphics[width=.95\textwidth]{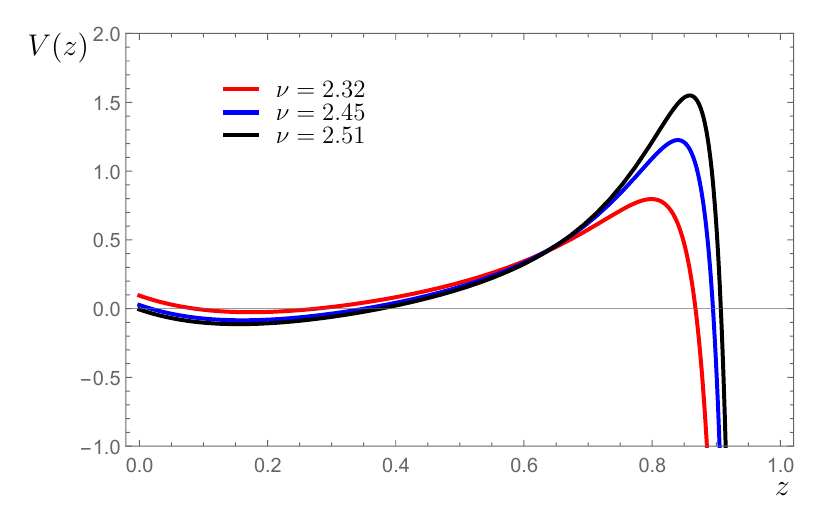}
		\caption{}
		\label{fig:Pot2a}
	\end{subfigure}
	\hfill     
	\begin{subfigure}[b]{.49\textwidth}
		\centering
		\includegraphics[width=.95\textwidth]{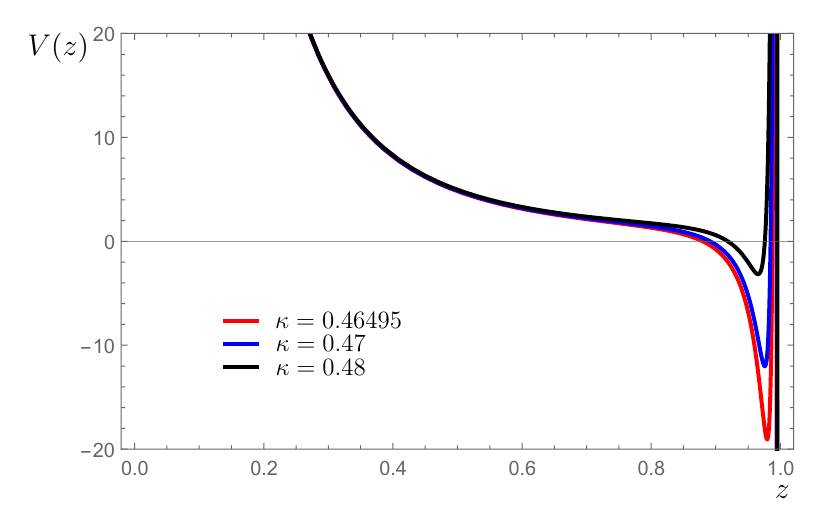}    
		\caption{}  
		\label{fig:Pot2b}
	\end{subfigure}
	\caption{The potential function as a function of holographic coordinate $z$. (a): For the specific value $m^{2}L^{2}=-2$ ( and taking $\kappa \approx 1.0242$) that renders the potential finite at the boundary. It can be seen that bound states can not exist in the near horizon region however, when reaching the AdS boundary, there is condensation slightly down below zero energy; representing the UV superconductor. (b): The potential function (\ref{potentialJi}) near extremality ($\nu=2.98$) for masses close to the AdS$_{2}$ bound ($m^{2}L^{2}=-0.24$) in which the scalar field condenses, representing the IR superconductor.}
    \label{fig:Vc3}
\end{figure}

A charged scalar field perturbation with energy $E$ propagating in the background (\ref{Metric3}), experiences the effects of the curvature and the black hole parameters due to the gauge-covariant derivative. Even more, the dependence of the potential (\ref{potentialJi}) on the scalar field mass is decisive in its behavior. Since the complete equation $|\psi|^{2}$ represents the density of Cooper pairs of the dual superconducting system, we look for bound states allowed by the potential in terms of the charges and horizon radius. If $z_{c}$ is the position at which the minimum of the potential (\ref{potentialJi}) occurs then, the neighborhood around it defines the bound states that we are looking for. To justify the use of the effective scalar equation (\ref{effectiveR}) in the next section, $z_{c}$ must happen close to the horizon therefore, we need to consider a fine-tuning of electromagnetic charges and $r_{+}$ in such a way, a local potential well allow for bound states (i.e., condensations), in the AdS$_{2}$ geometry. Although the potential (\ref{potentialJi}) is valid for all values of $(\mathcal{B},\mathcal{Q},r_{+})$, there is a strong restriction $\mathcal{F}<3r_{+}^{4}$ that arises from the positive definite character of the blackening factor $\chi$\footnote{Also, this restriction follows from the positive Hawking temperatures (\ref{hawking}) and more fundamentally, from the censorship condition (\ref{extremalC}).}. For that reason, we find it convenient to parameterize the charges as $\mathcal{B}\equiv\sqrt{\nu}\sin \kappa$ and $\mathcal{Q}\equiv\sqrt{\nu}\cos \kappa$. In this parametrization, $\nu\in\left(0,3\right)$ represents an \emph{extremality parameter} being $\nu=3$ the extremal case and $\nu=0$ the neutral black hole. The $\kappa$ parameter, on the other hand, controls the relative values between the electromagnetic charges, e.g., $\kappa=0$ corresponds to a Reissner-Nordström black hole without a magnetic charge. In this respect, refer to Figure \ref{fig:Pot1a}, in which, for specific values of $(\nu,\kappa)$, a particle with energy $E=0$ experiences the confining box of asymptotic AdS$_{4}$ and falls into the near horizon geometry where AdS$_{2}$ predominates. On the other hand, note that as $m^{2}L^{2}$ approaches $-2$ from above, bound states still exist in almost the entire bulk spacetime, as can be seen in Figure \ref{fig:Pot1b}.
Nevertheless, at exactly $m^{2}L^{2}= -2$, the potential becomes finite at the boundary and only in a narrow region near there, bound states arise with very small amplitude, see Figure \ref{fig:Pot2a}. This case corresponds (or is equivalent) to the UV Abelian-Higgs superconductor \cite{Hartnoll:2008vx} and it is what allowed us to obtain the thermal behavior of the upper critical magnetic field in subsection \ref{matchinProcedure}. Also, we can appreciate the appearance of a domain wall as we approach the horizon. On the other hand, close to extremality ($\nu\mapsto3$), with a magnetic charge slightly greater than the electric one, very robust and confining potential wells arise in the near horizon region, depicted in Figure \ref{fig:Pot2b}.

We end this report with some remarks about the conditions to have instabilities in the near extremal and near horizon geometry, in light of what was discussed above.

\subsection{SAdS$_{2}\times\mathbb{R}^{2}$ black hole and instabilities}\label{final}
\begin{figure}[ht]
	\centering
		\includegraphics[width=.6\textwidth]{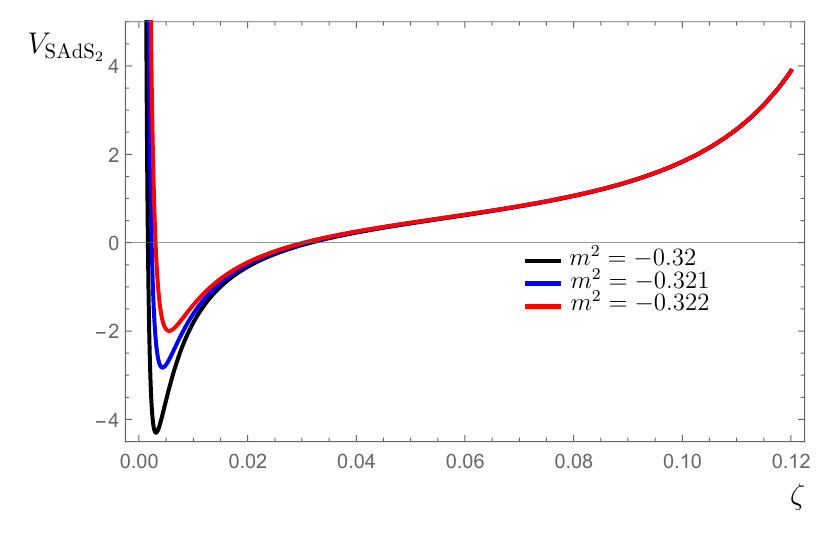}
         \caption{The Schrödinger potential for the scalar field equation (\ref{AdS2scalar}) as a function of $\zeta$ and different values of the bare mass in the window $(-9/4,-1/4)$. There are stable modes (in the sense of bound states described in this section) in the NENH geometry (\ref{AdS2bh}). According to (\ref{Hawkingfinal}), we can fix the temperature to $T=1$ for which the black hole horizon is located at $\zeta=1/2\pi\approx 0.16$ whereas the boundary is in $\zeta=0$. We have not been able to find bound states for $m^{2}L^{2}=-2$. That this value is stable in the UV but unstable in the IR, was previously reported in \cite{Iqbal:2010eh}.}
		\label{fig:AdSpot}
\end{figure}
The effective scalar field propagating in the extremal limit of the dyonic black hole (\ref{NHmetric}) can be unstable, according to (\ref{BFbounds}). This means that the conformal dimension of the scalar field becomes complex\footnote{The scalar field infinitely oscillates near the horizon for a sufficiently large value of black hole parameters.}. As was previously discussed, close to zero temperature, the dyonic solution also admits an AdS$_{2}\times\mathbb{R}^{2}$ Schwarzschild black brane structure (SAdS$_2$), having implications on the conformal dimension operator of the perturbative scalar field. An interesting question, formulated and solved for the first time in \cite{Gubser:2008px, Hartnoll:2008kx} is if the latter configuration supports an instability to form scalar hair. 
Here we provide a further exploration into this matter by analyzing the near extremal and near horizon geometry together with the associated effective scalar equation. We also construct the Schr\"{o}dinger potential of this effective equation in order to search for the existence of bound states indicating the presence of scalar hair.

Taking into account that the electromagnetic charges (\ref{Uinter}) have dimensions of length square, we can introduce a new parametrization in terms of a length scale $r_{*}$
\begin{equation}
    \mathcal{Q}=q r_{*}^2, \ \ \ \ \mathcal{B}=b r_{*}^2, \ \ \ \ \mathcal{F}=\left(q^2+b^2\right)r_{*}^4,
\end{equation}
with $q$ and $b$, dimensionless numbers. The horizon radius and extremality condition now read
\begin{equation}
    r_{+}=\left(\dfrac{q^2+b^2}{3}\right)^{1/4}\equiv \Tilde{r}, \qquad  T=0 \qquad \mbox{at} \qquad \dfrac{q^2+b^2}{3}\dfrac{r_{*}^4}{r_{+}^4}=1.
\end{equation}
Furthermore, considering the coordinates \cite{Faulkner:2009wj}
\begin{equation}\label{NearCoord}
   \mbox{NE:}\quad r-\Tilde{r}=\dfrac{\theta L_{eff}^{2}}{\zeta}\ll 1, \qquad \mbox{NH:} \quad r_{+}-\Tilde{r}=\dfrac{\theta L_{eff}^{2}}{\zeta_{0}}\ll 1,
\end{equation}
where NE and NH stand for near extremal and near horizon respectively. The above expressions capture the fact that the AdS$_2$ has a scaling limit $t\mapsto\theta^{-1}\tau$ and, as long we take $\theta\mapsto 0$ with $\zeta$ and $\tau$ finite, we can bring the full dyonic solution (\ref{metricCoef}) into a NENH metric
\begin{equation}\label{AdS2bh}
    ds^{2}=\dfrac{L_{eff}^{2}}{\zeta^{2}} \left(-f(\zeta)d\tau^{2}+\dfrac{d\zeta^{2}}{f(\zeta)}   \right)+\dfrac{\Tilde{r}^{2}}{L^{2}} d\vec{x}^{2},
\end{equation}
with blackening factor $f(\zeta)$ and Hawking temperature 
\begin{equation}\label{HawkingSAdS2}
    f(\zeta)=1-\dfrac{\zeta^{2}}{\zeta_{0}^{2}}, \qquad T=\dfrac{1}{2 \pi \zeta_{0}}.
\end{equation}
In this geometry, the horizon is located at $\zeta=\zeta_{0}$ while the boundary is at $\zeta=0$.

With this information at hand, the decoupled scalar field equation (\ref{Req1}), takes the form\footnote{Also, due to covariant derivative, with the coordinate (\ref{NearCoord}) we need to consider the transformed expressions of the gauge potential components $A_{\tau}$ and $A_{x_{2}}$ and take the limit $\theta\mapsto 0$. }
\begin{equation}\label{AdS2scalar}
    R''(\zeta)+\frac{f'(\zeta)}{f(\zeta)}R'(\zeta)+\dfrac{R(\zeta)}{\zeta^2 f(\zeta)}\left(\frac{q^2}{3\left(q^2+b^2\right)}\dfrac{\left(1-\zeta/\zeta_{0}\right)^2}{f(\zeta)}-L_{eff}^2 m^2-\left(\frac{3}{q^2+b^2}\right)^{1/2}\frac{b\lambda_{n}}{3}\right)=0.
\end{equation}
According to this equation, in the AdS$_{2}$ boundary $\zeta=0$, the asymptotic solution  of $R(\zeta)$ adopts the form
\begin{equation}
    R(\zeta\mapsto 0)\approx c_{1}\zeta^{\frac{1}{2}\left(1+\delta\right)}+c_{2}\zeta^{\frac{1}{2}\left(1-\delta\right)},
\end{equation}
being 
\begin{equation}
    \delta=\sqrt{1+4L_{eff}^2m^2+\dfrac{4b\lambda_{n}}{\sqrt{3\left(q^2+b^2\right)}}}.
\end{equation}
Hence, we can define an \emph{effective mass}
\begin{equation}
    M_{n}^2\equiv 4m^2+\dfrac{4b\lambda_{n}}{L_{eff}^2\sqrt{3\left(q^2+b^2\right)}},
\end{equation}
acquiring discrete labels due to the Landau Levels associated with (\ref{gammafunction}). In this sense, the critical magnetic field that allows stable modes in the SAdS$_{2}$ boundary needs to verify
\begin{equation}
    \vert b_{c}\vert= \frac{q \left(4 L_{eff}^2 m^2+1\right)}{\sqrt{\frac{16 \lambda_{n} ^2}{3}-\left(4 L_{eff}^2 m^2+1\right)^2}}\geq 0.
\end{equation}
In the BF window $(-9/4,-1/4)$ between AdS$_{4}$ and AdS$_{2}$ we can search for bound states of the scalar field equation. To achieve this, consider the Hawking temperature (\ref{HawkingSAdS2}), to express the blackening factor as
\begin{equation}\label{Hawkingfinal}
    f(\zeta)=1-\left(2 \pi T \zeta\right)^2.
\end{equation}
In the NENH geometry, the Schrödinger potential associated with the  equation (\ref{AdS2scalar}) reads
\begin{align}\label{Ads2potEq}
         V_{\text{SAdS}_{2}}(\zeta) &=  \dfrac{1}{3 \left(b^2+q^2\right) \zeta^2\left(1 -4 \pi ^2 \zeta ^2 T^2\right)^2}
         \Biggl(q^2 - 3 L_{eff}^2 m^2 (b^2 + q^2) - 4 \pi q^2 T \zeta\nonumber\\  \nonumber \\
          &+4 \pi^2 (3 b^2 (1 + L_{eff}^2 m^2) + (4 + 3 L_{eff}^2 m^2) q^2) T^2 \zeta^2\nonumber \\ \nonumber\\
         &+ \sqrt{3\left(q^2+b^2\right)}\left(4\pi^2 T^2 \zeta^2-1\right)b\Biggr),
\end{align}
where we choose $\lambda_{0}=1$ which corresponds to the more likely to condense mode. For some values of the bare mass $m^2$ and $(q,b)$, the potential, depicted in Figure \ref{fig:AdSpot}, shows bound states close to the boundary of SAdS$_2$. This is in agreement with the previous results obtained when considering the full scalar field equation, showing us that bound states can exist in the IR region. Such a result allows us to confirm the existence of superconductors in the IR regime, compatible with expectations from seminal works \cite{Gubser:2008px}.

\section{Discussion and concluding remarks}

In this report, we have studied a holographic model for a type-II superconductor defined on a (3+1)-dimensional black hole configuration with both, electric and magnetic charge, and which considers a perturbative charged scalar field as responsible for incorporating the superconducting nature of the background. 
In this sense, starting with the Einstein-Maxwell-Scalar theory, a suitable decoupling limit allowed us to consider a perturbative charged scalar field possessing information about the Einstein-Maxwell background throughout covariant derivatives. With a London gauge ansatz for the Maxwell field, the system was able to capture the behavior of the upper critical magnetic field as a function of the temperature, in the limit when the black hole electromagnetic charges are small compared with the Hawking temperature. We presented explicit expressions of this critical magnetic field in both, canonical and grand canonical ensemble, i.e., with fixed charge density and chemical potential, respectively. 
Regarding the description within the canonical ensemble, it is worth noting that our findings align with those predicted by the Ginzburg-Landau theory and other holographic advancements concerning type-II superconductors. Particularly, here we observe a linear behavior of the critical magnetic field near the transition temperature. On the other hand, concerning the description in the grand canonical ensemble, an interesting finding within this result is the nonlinear behavior of the curve near the critical temperature. We argue that this effect arises as a consequence of the difficulty introduced by keeping the chemical potential fixed instead of the density of charge.  
On the other hand, it is important to mention that our model falls short in capturing the expected thermal behavior of the order parameter, which represents the condensation of superconducting electrons. This is a limitation arising due to the considered nonbackreacting limit of the scalar field on the Maxwell field. We tackled out this limitation by considering the Schrödinger potential associated with the radial part of the scalar equation, revealing traces of condensation in virtue of the formation of local potential wells as a function of the holographic coordinate. In the neighborhood around a local minimum of such potential wells, bound states can exist, and we interpret these regions as locations in which condensation is likely to occur. It was found that, by considering the mass of the scalar field as $m^2L^2 = -2$, bound states exist near the boundary. This is what defines a UV superconductor, according to the preceding argument. Furthermore, through a suitable parametrization of black hole charges close to extremality, we were able to find bound states near the horizon, thus defining an IR superconductor.

Finally, to reinforce what was just described in the preceding lines, we explored the near extremality and near horizon geometry, adopting a Schwarzschild AdS$_{2}\times\mathbb{R}^{2}$ structure, obtaining the effective scalar field equation. We found the corresponding Breitenlohner-Freedman bound (which acquired contributions of the black hole charges). The Schrödinger potential for this effective scalar equation also captures potential wells in a fine-tuning range of black hole parameters. We confirmed that the scalar field is not stable in the IR when $m^2L^2=-2$, nonetheless, a fine-tuning of the black hole parameters gives rise to stable modes when $m^2 L_{eff}^2$ is around $-1/3$. We believe that the existence of such bound states in the IR regime promotes the exploration of this theory beyond the present work. For instance: In the lattice structure in the IR regime in the context of Abelian-Higgs theory, the renormalization group flows from the UV to the IR for relevant operators in the context of the potential wells described here \cite{Faulkner:2010jy}; and the effects due to rotational black hole parameters with Lifshitz scaling (as in \cite{Herrera-Aguilar:2021top}) in the near horizon geometry.

The IR structure of AdS in Einstein-Maxwell theory has been seriously investigated in a series of papers \cite{Faulkner:2009wj, Hartnoll:2020fhc, Hartnoll:2016apf, Faulkner:2010gj,Faulkner:2010jy}, focusing on fermionic degrees of freedom and the holographic Fermi surface in the quantum criticality state of matter. These fascinating works constitute good guidance to explore and extend the bosonic perspective of superconductivity. Nevertheless, the approaches we want to address have not been reported hence, constitute a motivation for future research.

\section*{Acknowledgments}
\begin{small}
All authors have benefited from the CONAHCYT grant No. A1-S-38041, while RCF and AHA were supported by the VIEP-BUAP grant No. 122. 
MCL thanks the financial assistance provided by a CONAHCYT postdoctoral grant No. 30563.
JAHM acknowledges support from CONAHCYT through a PhD grant No. 750974. All authors are grateful to Mehrab Momenia and Uriel Noriega Cornelio for enriching discussions.
\end{small}

\bibliography{manuscript_AdSlimit}
\bibliographystyle{./utphys}

\end{document}